\documentclass[aps, longbibliography, amsmath,amssymb, superscriptaddress, 12pt]{revtex4-2}
\usepackage[normalem]{ulem}
\usepackage{color,graphicx,subfigure}
\usepackage{textcomp}
\usepackage[]{hyperref}
\hypersetup{colorlinks=true,linkcolor=blue,citecolor=blue,urlcolor=blue,pdfpagemode=UseNone}
\usepackage{xcolor}
\usepackage{floatrow}
\usepackage{floatflt}

\newcommand{\bscco}{$\textrm{Bi}_2\textrm{Sr}_2\textrm{Ca}\textrm{Cu}_2\textrm{O}_{8+\delta}$}
\newcommand{\hgcuprate}{HgBa$_{2}$CuO$_{4+\delta}$}
\newcommand{\BSCO}{Bi$_2$Sr$_{2}$CuO$_{6+\delta}$}
\newcommand{\ybco}{$\textrm{Y}\textrm{Ba}_2\textrm{Cu}_3\textrm{O}_{6+\delta}$}
\newcommand{\ncco}{$\textrm{Nd}_{2-x}\textrm{Ce}_x\textrm{Cu}\textrm{O}_{4}$}
\newcommand{\lcco}{$\textrm{La}_{2-x}\textrm{Ce}_x\textrm{Cu}\textrm{O}_{4}$}
\newcommand{\lbco}{$\textrm{La}_{2-x}\textrm{Ba}_x\textrm{Cu}\textrm{O}_{4}$}
\newcommand{\lsco}{$\textrm{La}_{2-x}\textrm{Sr}_x\textrm{Cu}\textrm{O}_{4}$}

\begin{document}

\title[Dynamic Charge Order from Strong Correlations in the Cuprates]{Dynamic Charge Order from Strong Correlations in the Cuprates} 

\author{Eduardo H. da Silva Neto}
\affiliation{\footnotesize \mbox{Department of Physics, Yale University, New Haven, Connecticut, USA}}
\affiliation{\footnotesize \mbox{Energy Sciences Institute, Yale University, West Haven, Connecticut, USA}}
\affiliation{\footnotesize \mbox{Department of Applied Physics, Yale University, West Haven, Connecticut, USA}}

\author{Alex Frano}
\affiliation{\footnotesize \mbox{Department of Physics, University of California San Diego, La Jolla, CA, United States}}

\author{Fabio Boschini}
\affiliation{\footnotesize Centre \'Energie Mat\'eriaux T\'el\'ecommunications, Institut National de la Recherche Scientifique, Varennes, QC, Canada}

\begin{abstract}

Charge order has been a central focus in the study of cuprate high-temperature superconductors {due to its intriguing yet not fully understood connection to superconductivity}. Recent advances in resonant inelastic x-ray scattering (RIXS) in the soft x-ray regime have enabled the first momentum-resolved studies of \textit{dynamic} charge order correlations in the cuprates. This progress has opened a window for {a more nuanced} investigation into the mechanisms behind the formation of charge order (CO) correlations. This review provides an overview of RIXS-based measurements of dynamic CO correlations in various cuprate materials. It specifically focuses on electron-doped cuprates and Bi-based hole-doped cuprates, where the CO-related RIXS signals may reveal signatures of the effective Coulomb interactions. {This aims to explore a connection between two central phenomena in the cuprates: strong Coulomb correlations and CO-forming tendencies.}
{Finally, we discuss current open questions and potential directions for future RIXS studies as the technique continues to improve and mature, along with other probes of dynamic correlations that would provide a more comprehensive picture.}
\end{abstract}

\maketitle

\section{Introduction \label{sec:intro}}

Cuprate high-temperature superconductors exhibit a variety of emergent phenomena, including antiferromagnetism, superconductivity, charge order, and strange metal phases. Charge order (CO) has been a central focus in the study of cuprates for a long time, with the initial detection of stripe order in La-based cuprates by neutron scattering in 1995 and the first evidences for charge order in Bi-based cuprates coming from scanning tunneling microscopy and spectroscopy (STM/S) measurements \cite{Comin_review, Frano_2020_review, tranquada_evidence_1995, Hoffman_2002, Vershinin_2004, Howald_2003, Abbamonte_2005, Wise_2008, Ghiringhelli_CDW_2012, Chang_CDW_2012, Comin_CDW_2014, daSilvaNeto_CDW_2014, Tabis_2014, daSilvaNeto_Science_2015, daSilvaNeto_SciAdv_2016}.
A significant breakthrough occurred in 2011 when an incommensurate CO, independent of spin modulations, was detected in \ybco{} (YBCO) \cite{Wu_2011, Ghiringhelli_CDW_2012, Chang_CDW_2012}. {One of the original} momentum-resolved detections was obtained from resonant inelastic x-ray scattering experiments (RIXS), but in the same work it was shown that a CO peak in momentum space could be detected with energy-integrated resonant x-ray scattering (EI-RXS), where all the scattered photons are counted by a finite area detector without resolving their energy \cite{Ghiringhelli_CDW_2012}.
Soon after the discovery in YBCO, EI-RXS was used to detect CO in the major hole- and electron-doped cuprate families, including YBCO, \BSCO \,(Bi-2201), \bscco \,(Bi-2212), \hgcuprate \,(HgBCO), \ncco \,(NCCO), and \lcco. \cite{Comin_review, Frano_2020_review, tranquada_evidence_1995, Abbamonte_2005, Ghiringhelli_CDW_2012, Chang_CDW_2012, Comin_CDW_2014, daSilvaNeto_CDW_2014, Tabis_2014, daSilvaNeto_Science_2015, daSilvaNeto_SciAdv_2016}.
Although the original detection in YBCO was performed using RIXS, EI-RXS was chosen for many studies as it allowed the CO to be characterized as a function of temperature, magnetic fields, and doping in a more time efficient manner. However, this approach could not distinguish whether the scattering was elastic (related to static charge modulations) or inelastic (related to dynamic charge correlations). 
In the last seven years, technical advances in soft x-ray RIXS improved the energy resolution at the Cu-$L_3$ edge from approximately $130$\,meV to less than $40$\,meV \cite{BROOKES_ID32_instrument, 2ID_instrument, SIX_instrument, Zhou:rv5159}, enabling explorations of dynamic electron correlations near the CO wavevector \cite{Arpaia_review, Chaix2017, daSilvaNeto_PRB_2018, Arpaia_Science, BYu_Hg2019, Miao_2017, Miao_PRX_2019,Li_PNAS_2020, lee2021melting, boschini2021dynamic, Huang_PRX_2021, Lu_PRB_2022_phonons_2212, Scott_SciAdv_2023, Arpaia2023}. {These measurements have shown that indeed the charge order is not only static, but shows distinct features in the inelastic spectrum. To date}, a universal picture of dynamic CO correlations has yet to be established.

Charge order has been observed as short-range, incommensurate correlations, leading to proposals that it can be understood in terms of a Fermi surface instability or a nesting condition \cite{Comin_review}. This perspective has led many to view CO as a minor peculiarity of the normal state of underdoped cuprates, essentially an accident of the Fermi surface. However, this early astute interpretation lacked the insights into inelastic dynamical charge order correlations that later emerged with the full exploitation of the energy-resolving capability of RIXS measurements. 
Another view sees CO as a consequence of the strong electron correlations that are at the heart of the cuprate problem \cite{MACHIDA1989192, zaanen_charged_1989, Emery_1993, Huang1161, Zheng1155}. {Given the central role of strong correlations in cuprate physics, we hypothesize, based on the evidence reviewed here, that
they should be considered in any attempt to understand dynamic CO correlations.} 

Indeed, much of the interest in CO arose from early theoretical proposals suggesting that the formation of charge or stripe order is a natural way for the doped Mott insulator to frustrate the tendency toward phase separation \cite{Emery_1993}. While short-range local Hubbard-like interactions would cause spatial phase separation into metallic and insulating regions, long-range Coulomb interactions prevent this long-wavelength phase separation and result in short-wavelength charge fluctuations instead. These fluctuations may manifest as either dynamic CO modes or incommensurate static CO, or both. Despite numerous EI-RXS and RIXS studies on CO in cuprates, connecting CO correlations to the effective Coulomb interactions has been challenging. This review will explore prominent experiments where Cu-$L_3$ RIXS measurements provide insights into {how effective Coulomb interactions may relate to CO}. 

This review does not aim to comprehensively cover the phenomenon of charge ordering in cuprates. Several aspects, such as the CO doping dependence, differences and similarities between cuprate families, and the determination of the CO intra-unit-cell structure, are beyond its scope; for these topics, readers are directed to existing literature \cite{Comin_review, Frano_2020_review, Arpaia_review}. Instead, this focused review delves specifically into how Coulomb interactions may relate to CO. Additionally, our understanding of CO in the cuprates has evolved significantly over the past 13 years, closely tied to advancements in EI-RXS and RIXS techniques in the soft x-rays. As these techniques improve, they not only expand our knowledge but also reveal new questions and expose current experimental limitations. Therefore, this review also highlights important technical developments, discusses our present limitations, and suggests possible future directions.

\section{Discussion}

\subsection{The cuprate Cu-$L_3$ RIXS spectrum}

\begin{figure}[t]
\centering
\includegraphics[scale=0.8]{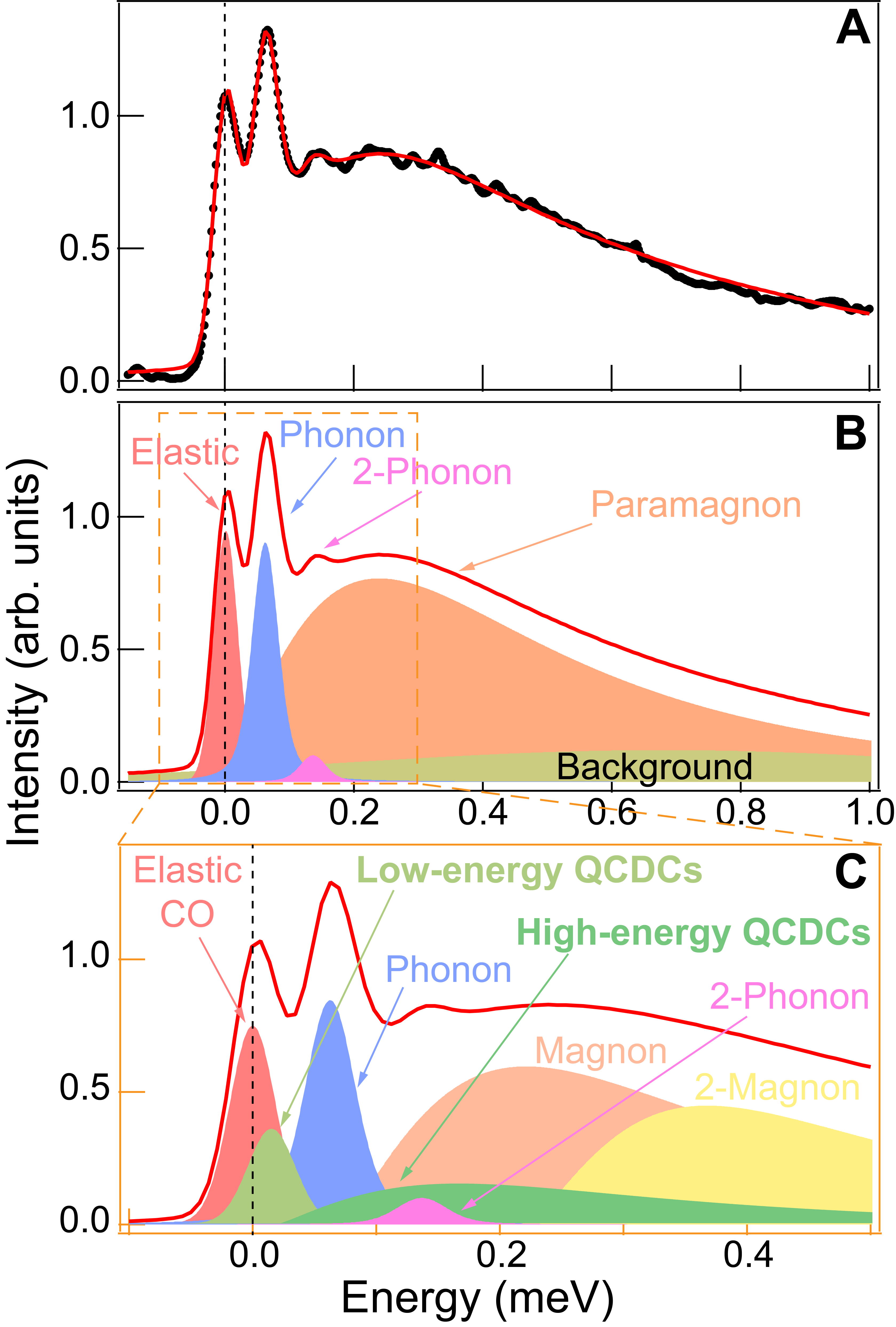}
\caption{
\footnotesize 
Cuprate Cu-$L_3$ RIXS spectrum. (\textbf{A}) Example of a RIXS spectrum at $q=0.27$\,\textit{r.l.u.} for $\phi=0^\circ$. The red line is a fit to the spectrum. (\textbf{B}) Contributions to the fit of (A): a quasi-elastic peak, a phonon peak, a two-phonon peak, a broad paramagnon, and a broad background. (\textbf{C}) Sketch of other possible contributions to the RIXS spectrum of (A): low- and high-energy QCDCs, as well as a magnon and bimagnon peak instead of the broad paramagnon feature.
Figures (A, B) are adapted from (Scott et al., 2023) \cite{Scott_SciAdv_2023}. Reprinted with permission from AAAS. 
\label{fig:0}}
\end{figure}

The Cu-$L_3$ RIXS cross-section in cuprates is extremely rich. It is sensitive to charge order correlations \cite{Comin_review, Frano_2020_review, tranquada_evidence_1995, Hoffman_2002, Vershinin_2004, Howald_2003, Abbamonte_2005, Wise_2008, Ghiringhelli_CDW_2012, Chang_CDW_2012, Comin_CDW_2014, daSilvaNeto_CDW_2014, Tabis_2014, daSilvaNeto_Science_2015, daSilvaNeto_SciAdv_2016, Arpaia_review, Chaix2017, daSilvaNeto_PRB_2018, Arpaia_Science, BYu_Hg2019, Miao_2017, Miao_PRX_2019,Li_PNAS_2020, lee2021melting, boschini2021dynamic, Huang_PRX_2021, Lu_PRB_2022_phonons_2212, Scott_SciAdv_2023, Arpaia2023}, phonons (including studies of electron-phonon coupling and two-phonon processes) \cite{Chaix2017, Rossi_PRL_EPC, Lin_PRL_2020,Peng_PRL_2020,Li_PNAS_2020,Wang_SciAdv_2021,lee2021melting, Lu_PRB_2022_phonons_2212, Giacomo_PRR_phonon_flat, Huang_PRX_2021, YY_Peng_PRB_2022_EPC, Scott_SciAdv_2023, Scott_PRB_2024}, superconducting and pseudo gaps \cite{Suzuki_SCGap_2018, Merzoni_PRB_2024}, magnetic excitations (magnon and bi-magnon excitations) \cite{Hill_PhysRevLett.100.097001, Braicovich_PRL_2010_Magnons, LeTacon2011, Lee2014, Ishii_Soin_2014, YY_Peng_2015, MM_PRL_2015, Fumagalli_NBCO_pol, Betto_PRB_2021_Multiple_magnon}, plasmons \cite{Hepting_Plasmon2018, Lin_npjQM_2020_plasmons, Nag_PRL_2020_Plasmons, Singh_PRB_2022_Plasmons, Hepting_PRL_2023_plasmons,  Hepting_PhysRevB2023_plasmon}, dd orbital excitations \cite{MMS_dd, Fumagalli_NBCO_pol}, and the charge transfer gap. Within this rich cross-section, static CO appears prominently as an enhancement of the elastic line at in-plane wavevectors along the Cu-O bond direction, i.e., $\mathbf{q}=(\pm q_{CO}, 0)$ and $\mathbf{q}=(0, \pm q_{CO})$. However, dynamic CO correlations do not appear as a salient feature of the Cu-$L_3$ RIXS energy-loss spectrum. 
In fact, an agnostic minimal model of the spectrum in the zero to $1$\,eV range, requires the following five components though not necessarily a dynamic CO feature: 
(\textit{i}) a quasi-elastic peak, (\textit{ii}) the bond-stretching phonon around $70$\,meV, (\textit{iii}) two-phonon mode (from the bond-stretching phonon) near $140$\,meV, (\textit{iv}) a damped harmonic oscillator representing spin-flip processes in the $100$ to $800$\,meV range, and (\textit{v}) a small broad background \cite{Scott_SciAdv_2023}. The origin of (\textit{v}) remains unknown, but some possibilities include scattering from extra charge carriers from doping, a Stoner continuum or high energy QCDCs, see Fig.\,\ref{fig:0}(A,B). Beyond those, great caution must be taken when attempting to add additional terms to the curve fitting functions, although other contributions can be inferred, as depicted in Fig.\,\ref{fig:0}(C). Indeed, subtler features relating to dynamic CO correlations can be detected by performing careful momentum mapping measurements as a function of temperature, as discussed in Sec.\,\ref{ssec: dynamic CO}. 
Furthermore, an important quantity that can significantly aid in separating overlapping signals in the spectrum is the polarization of the outgoing photon, which provide information to distinguish between charge and spin scattering. This is accomplished in polarimetric RIXS (pol-RIXS), where the scattered photons are analyzed not only in terms of their energy loss but also in terms of their polarization \cite{Braicovich, MM_PRL_2015, Fumagalli_NBCO_pol, Betto_PRB_2021_Multiple_magnon, daSilvaNeto_PRB_2018, Hepting_Plasmon2018, Scott_PRB_2024}. 
For example, pol-RIXS measurements were used to demonstrate the two-phonon origin of the feature near $140$\,meV, showing that it appears in the non-cross polarized scattering channel, $\sigma$-$\sigma'$, but not in the cross-polarized channel, $\sigma$-$\pi'$, where the unprimed (primed) symbol refers to the incoming (scattered) photon polarization \cite{Scott_PRB_2024}. In another example, pol-RIXS measurements show that the damped harmonic oscillator feature, often simply called a `paramagnon' feature, is actually composed of one- and two-magnon processes \cite{Betto_PRB_2021_Multiple_magnon}. 
Despite being a very powerful technique {which can resolve fundamental aspects of the scattered signals}, the current state-of-the-art pol-RIXS technology is inefficient, typically requiring ten times longer acquisition times than standard RIXS. 
These lengthy acquisition times have led scientists to focus more on standard RIXS measurements rather than pol-RIXS. However, even with these current limitations, many potential pol-RIXS experiments in cuprates and other materials are feasible with existing technology, indicating potential for a wider exploration of pol-RIXS studies. Moreover, as the method becomes more efficient, it will enable new experiments that will undoubtedly uncover important information.

\subsection{Dynamic electron correlations near the charge order wavevector \label{ssec: dynamic CO}}

Given the complexity of the RIXS signal discussed above, how can we detect dynamic  CO correlations within such a rich spectrum? This is typically accomplished through careful momentum mapping which reveal subtle dynamic features near the charge order wavevector, $\mathbf{q}_{CO}$. These features can be broadly categorized into low-energy excitations, below the bond-stretching phonon energy (i.e., $E < 70$\,meV), and high-energy excitations, spanning the 150 to 700\,meV range. Low-energy features, existing within the same energy range as various phonon branches, have been identified in Bi-2212 \cite{Chaix2017, lee2021melting, Lu_PRB_2022_phonons_2212, Scott_SciAdv_2023}, Bi-2201 \cite{Li_PNAS_2020}, YBCO \cite{Arpaia_Science, Arpaia2023}, HgBCO \cite{BYu_Hg2019}, \lsco \cite{Huang_PRX_2021}, and \lbco \cite{Miao_PRX_2019}, with experimental methodologies varying between studies due to the peculiarities of different materials. 
For instance, in Bi-based materials, the presence of low-energy CO correlations is inferred through the observation of a softening of the RIXS-measured bond-stretching phonon \cite{Chaix2017, lee2021melting, Lu_PRB_2022_phonons_2212, Scott_SciAdv_2023}, while in YBCO, the detection of a similar low-energy CO mode was achieved through the careful subtraction of temperature-dependent data \cite{Arpaia_Science, Arpaia2023}. 
High-energy features, observed in NCCO \cite{daSilvaNeto_PRB_2018}, HgBCO \cite{BYu_Hg2019}, and Bi-2212 \cite{boschini2021dynamic, Scott_SciAdv_2023}, overlap with paramagnon excitations and are broader. They are usually detected by examining the momentum-dependence of the RIXS spectrum integrated over an energy window or through meticulous temperature-dependent measurements and background subtraction.

\begin{figure}[htbp]
\begin{center}
\includegraphics[width=\linewidth]{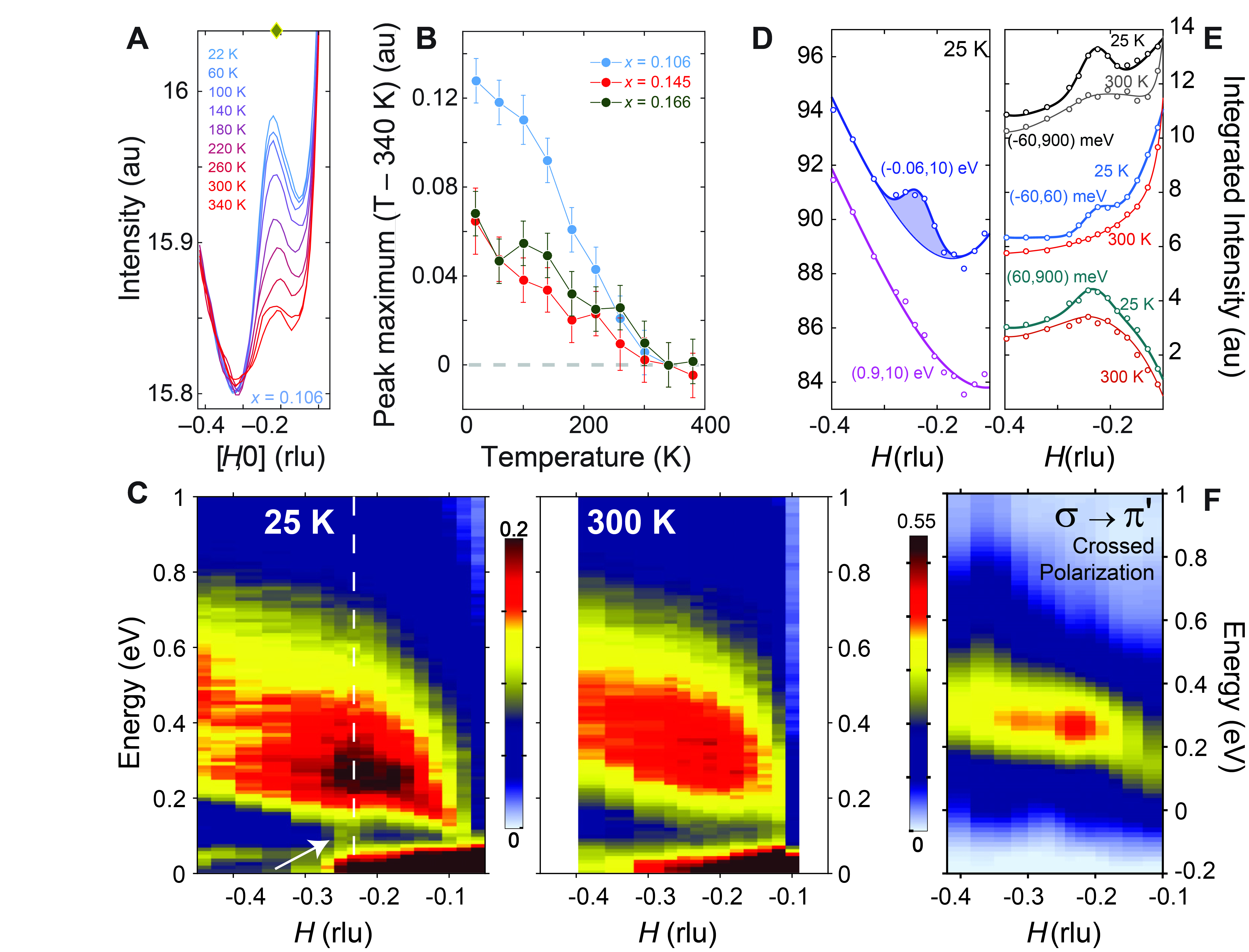}
\end{center}
\caption{
\footnotesize CO correlations in NCCO. (\textbf{A}) Temperature dependence of EI-RXS momentum scans showing the presence of a peak at $q_{CO}$, marked by the yellow diamond. (\textbf{B}) Temperature dependence of the CO peak maximum after subtraction of the $340$\,K peak maximum, for NCCO at different doping values. (\textbf{C}) Energy-momentum structure of the excitations in NCCO measured with $\sigma$ scattering at $25$ and $300$\,K. The dashed line in marks $q_{CO}$ from the energy-integrated data in
panel A. (\textbf{D}) 25\,K RIXS signal integrated over different energy
ranges showing that the peak at $q_{CO}$ originates from energies below
$900$\,meV . (\textbf{E}) 25\,K RIXS over different energy integration windows for $25$ and $300$\,K. (\textbf{F}) Energy-momentum structure in the $\sigma$-$\pi '$ channel, which is primarily composed of single spin-flip processes.  The measurements shown in Figs. (\textbf{A},\textbf{C}-\textbf{F}), were done on a non-superconducting NCCO $x=0.106$ sample. 
Figures (\textbf{A},\textbf{B}) are from (da Silva Neto et al., 2016) \cite{daSilvaNeto_SciAdv_2016}. Reprinted with permission from AAAS. Figures (\textbf{C}-\textbf{F}) reprinted with permission from (da Silva Neto et al., 2018) \cite{daSilvaNeto_PRB_2018}. Copyright 2018 by the American Physics Society.
\label{fig:1}}
\end{figure}

In exploring the relationship between strong electron correlations and CO, high-energy CO features offer valuable insights. These higher energy scales, comparable to the magnetic exchange coupling $J$, are more compatible with strong correlation physics. 
Perhaps, the clearest realization of the high-energy signal is observed in the electron-doped cuprate NCCO, as discussed in the following section.

\subsection{High energy dynamic CO correlations, the case of NCCO}

The first indications of CO in electron-doped cuprates came from EI-RXS measurements in NCCO \cite{daSilvaNeto_Science_2015}. The peak at $\mathbf{q}_{CO}$ decreases in intensity with increasing temperature, but, up to $420$\,K, an onset temperature is not clear in the EI-RXS measurements \cite{daSilvaNeto_Science_2015, daSilvaNeto_SciAdv_2016}, Figure \ref{fig:1}(A,B). 
This linear temperature dependence, also observed in Bi-2212 \cite{boschini2021dynamic}, contrasts with typical mean-field-like order parameter behavior. It suggests that dynamic correlations, expected to persist at very high temperatures, significantly contribute to the peak spectral width in the EI-RXS measurements. 
To analyze the different static and dynamic contributions to this peak, RIXS measurements were employed \cite{daSilvaNeto_PRB_2018}. It was observed that approximately $50$\% of the low-temperature peak in EI-RXS is composed of dynamic CO correlations, clearly observed in the energy-momentum maps at $\mathbf{q}_{CO}$ in the same energy range as the paramagnon excitations, Figure \ref{fig:1}(C-E). At $300$\,K, the elastic peak disappears, but dynamic CO correlations remain, Figure \ref{fig:1}(C,E). Given the overlap between the high-energy feature at $\mathbf{q}_{CO}$ and the dispersive paramagnon feature, a natural question arose about whether the scattering was spin-flip or non-spin-flip in nature. The use of pol-RIXS revealed that the enhancement was primarily in the $\sigma$-$\pi'$ channel (cross-polarization) but not in the $\sigma$-$\sigma'$ (non-crossed) channel, indicating a spin-flip scattering process \cite{daSilvaNeto_PRB_2018}, Figure \ref{fig:1}(F).

At first inspection, the experimental results on NCCO present a puzzle. On one hand, the elastic and inelastic signals appear to be linked since they occur at the same $\mathbf{q}_{CO}$ wavevector, which was confirmed for two different dopings, each with distinct $\mathbf{q}_{CO}$. On the other hand, the elastic and inelastic signals appear in distinct polarization channels, $\sigma$-$\sigma'$ and $\sigma$-$\pi'$, respectively. However, a resolution was proposed in \cite{daSilvaNeto_PRB_2018}, by noting that that temporal fluctuations of a CO pattern on top of underlying antiferromagnetic correlations require the transfer of charge between neighboring sites with alternating spins. Therefore, some dynamic charge order processes will necessarily involve spin flips.

\begin{figure}[htbp]
\begin{center}
\includegraphics[width=\linewidth]{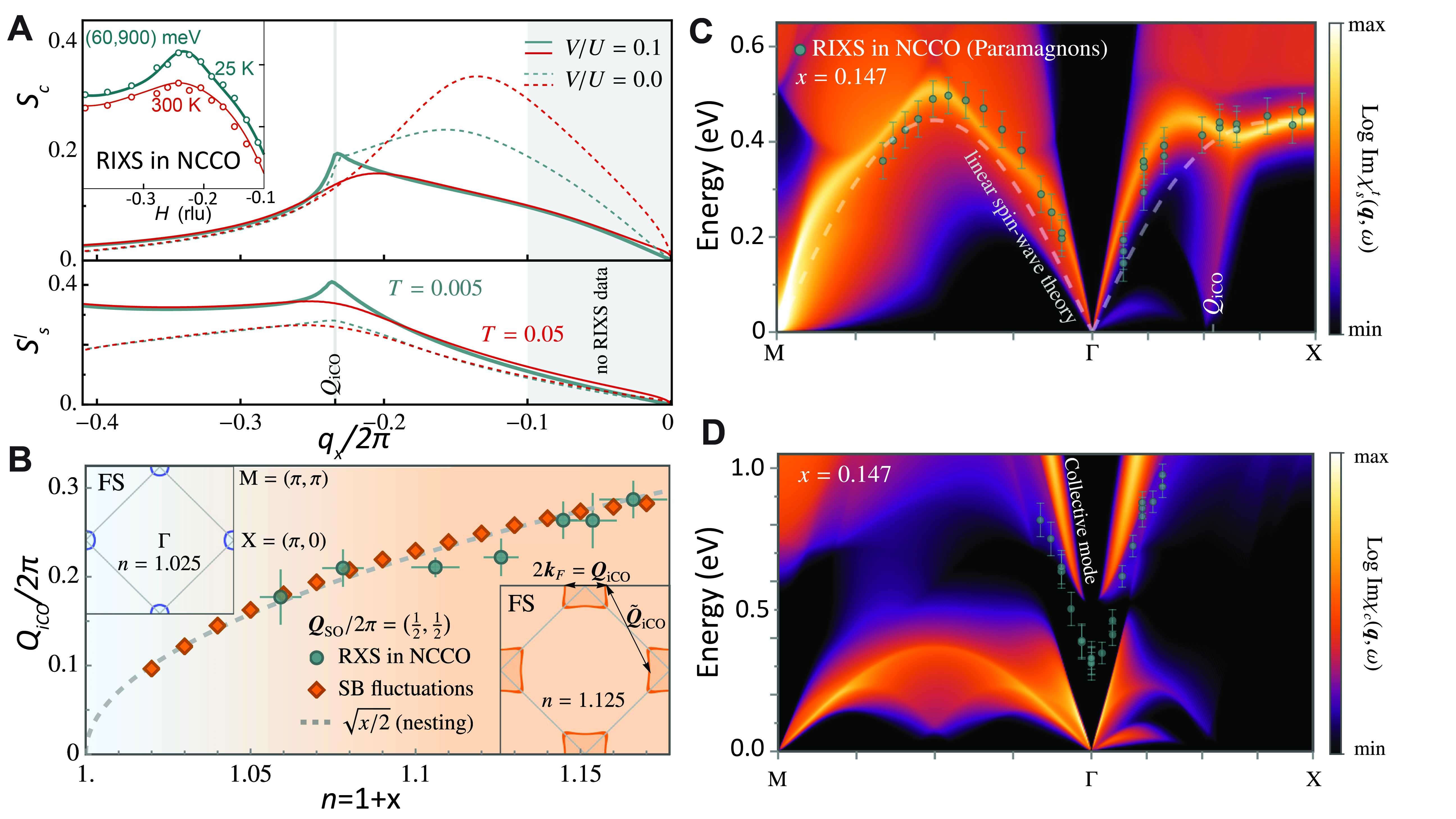}
\end{center}
\caption{
\footnotesize
Theory of charge and spins susceptibility in electron-doped cuprates and comparison to RIXS measurements in NCCO. (\textbf{A}) Charge ($S_c$) and longitudinal ($S^l_s$) spin-structure factor for $x = 0.108$ NCCO as a function of momentum integrated over energy ranges similar to the experiments in \cite{daSilvaNeto_PRB_2018} (Fig.\,\ref{fig:1}E and inset). (\textbf{B}) $q_{CO}$ wavevector inferred from the calculations and compared to EI-RXS data in NCCO \cite{daSilvaNeto_SciAdv_2016}. (\textbf{C}) Paramagnon dispersion along the high-symmetry directions, which appears as bright lines in the color-coded plot of the transverse spin susceptibility along with RIXS data in NCCO (circles with error bars) obtained from \cite{Lee2014}. (\textbf{D}) Gapped collective mode around the $\Gamma$-point in the charge susceptibility $\chi_c$ along with RIXS data in NCCO from \cite{Lee2014} displayed by green circles with error bars. 
Reprinted with permission from (Riegler et al., 2023) \cite{Riegler_PRB_2023}. Copyright 2023 American Physics Society.
\label{fig:2}}
\end{figure}

The existence of dynamic correlations at the $\sigma$-$\pi'$ channel at the same energy scale as $J$ indicates a connection between strong correlation physics and charge order formation. With this rich and detailed phenomenology available, one may ask: what is the simplest Hamiltonian that describes this combined phenomenology? What is the effective Coulomb interaction in that Hamiltonian? 
A recent theoretical work based on an extended one-band Hubbard model successfully reproduced various aspects of the RIXS experiments on NCCO, including paramagnon dispersion, a plasmonic mode, charge order, the doping dependence of $\mathbf{q}_{CO}$, and a significant dynamic CO signal at paramagnon energy scales \cite{Riegler_PRB_2023}, Figure \ref{fig:2}. While model calculations often have to be modified to improve agreement with an experimental aspect at the expense of another, the model in  \cite{Riegler_PRB_2023} shows excellent simultaneous agreement with all the above-mentioned experimental findings. A key ingredient necessary for this excellent agreement is the existence of a nearest-neighbor Coulomb repulsion $V$ in addition to the onsite repulsion $U$. Within this model, it was found that a moderate $V$ removes the propensity for phase separation with doping, leading to the formation of CO, and associated dynamic correlations, similar to the general mechanism discussed in Sec.\,\ref{sec:intro}.
Crucially, the comprehensive and precise experimental RIXS studies provide extensive constraints on the model, naturally increasing confidence in the effective Coulomb interaction obtained from it.

\subsection{Quasi-Circular Dynamic Correlations in Bi-2212} \label{subsec_HE_QCDCs}

To investigate the fingerprints of the effective Coulomb interaction in cuprates, studies of Bi-2212 adopted a new approach. Traditionally, RIXS experiments, constrained by limited synchrotron beam time, focus on the two high-symmetry directions of the 2D in-plane scattering Brillouin zone. However, the $q_x$-$q_y$ structure of the Cu-$L_3$ RIXS cross-section 
contains signatures of the effective Coulomb interaction, which we review in this section. Before that discussion, however, we will discuss two notable EI-RXS studies that underscore the significance of full in-plane RIXS mapping.

\begin{figure}
\begin{center}
\includegraphics[width=0.67\linewidth]{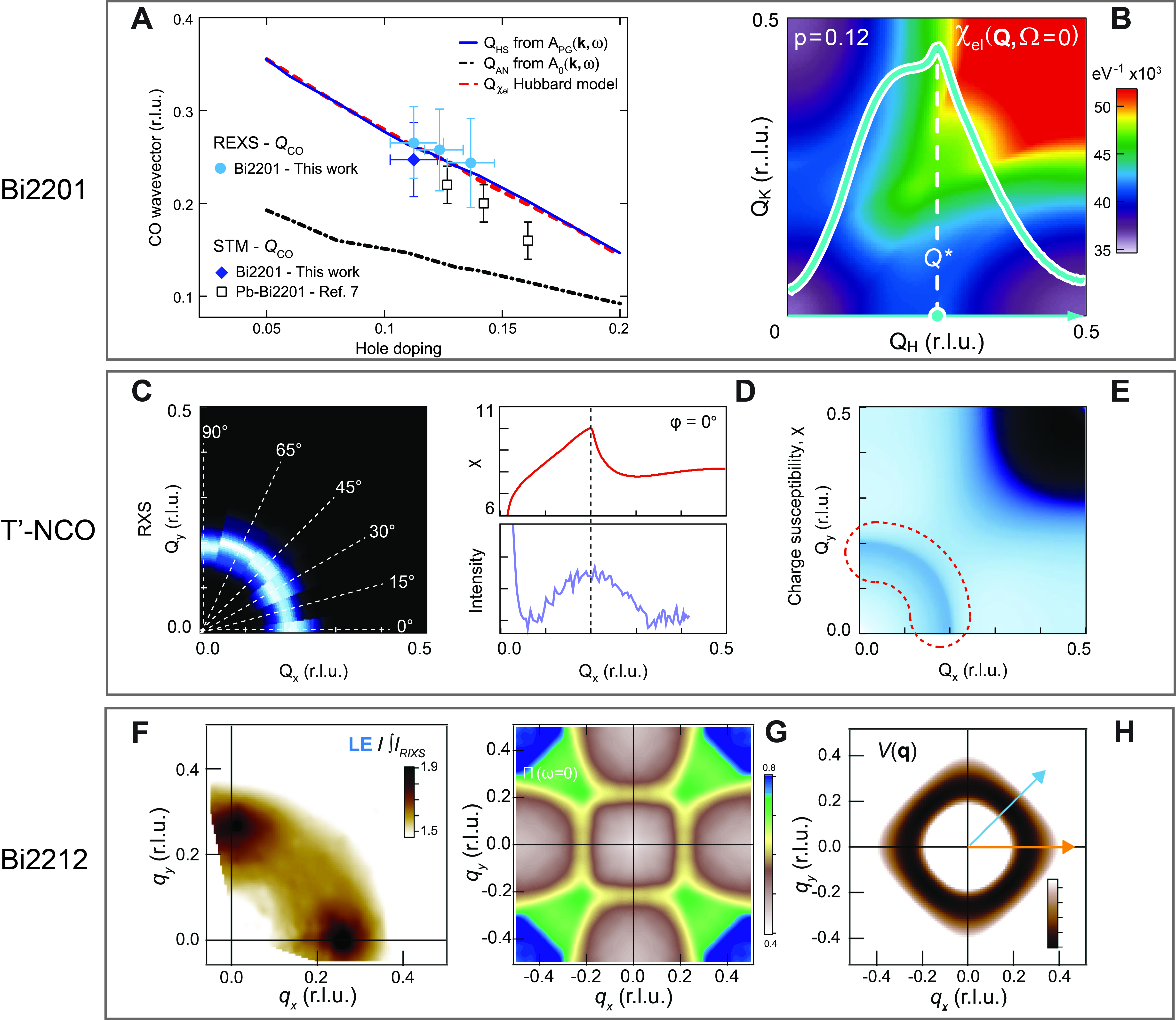}
\end{center}
\caption{
\footnotesize
(\textbf{A}) Doping dependence of $q_{CO}$ as determined by EI-RXS and STM on Bi2201 \cite{Comin_CDW_2014, Wise_2008}. Also shown are evolution of the Fermi surface-derived wavevectors $Q_{AN}$ (antinodal nesting) and $Q_{HS}$ (arc tips) measured from the ARPES spectral function, as well as the doping dependent wavevector $Q_{\chi _{el}}$ 
from the calculated Lindhard electronic susceptibility. (\textbf{B}) Static Lindhard susceptibility aclculated for $p=0.12$ doping. The cut along $Q_H$ (full blue line) is overlaid, and the local maximum at $Q_{\chi _{el}}$ 
$\approx 0.26$ is highlighted.
(\textbf{C}) EI-RXS measurements of $T'$-Nd$_2$CuO$_4$ over the $q_x$-$q_y$ plane. (\textbf{D}) Comparison of static Lindhard susceptibility and EI-RXS intensity along the $Q_x$ direction. (\textbf{E}) Static Lindhard susceptibility calculated over the full scattering plane. (\textbf{F}) RIXS structure in the $q_x$–$q_y$ plane integrated over [$-0.4,0.9$]\,eV (LE), normalized to the total fluorescence, i.e., the RIXS spectrum integrated in the [$-4,25$]\,eV energy range, at $50$\,K. 
(\textbf{G}) Static Lindhard susceptibility calculated for Bi-2212 over the $q_x$-$q_y$ plane. Similar to the calculations for Bi-2201 (\textbf{B}) and $T'$-Nd$_2$CuO$_4$ (\textbf{E}), it shows a maximum intensity along the diagonal direction ($q_x=q_y$). 
(\textbf{H}) The structure of the calculated Coulomb interaction $V_{\textrm{eff}}$ calculated using known parameters for Bi-2212.
Figures (\textbf{A},\textbf{B}) from (Comin et al., 2014) \cite{Comin_CDW_2014}. Reprinted with permission from AAAS. Figures (\textbf{C}-\textbf{E}) first published in (Kang et al., 2019) \cite{Kang2019} by Springer Nature. Figures (\textbf{F}-\textbf{H}) are reproduced from (Boschini et al., 2021) \cite{boschini2021dynamic}. Reference (Boschini et al., 2021) is an open access article licensed under a Creative Commons Attribution 4.0 International License.
\label{fig:3}}
\end{figure}

The first study compared EI-RXS measurements of the $q_{CO}$ in Bi-2201 as a function of doping to calculations of the static Lindhard susceptibility \cite{Comin_CDW_2014}. It was found that along the Cu-O bond direction, the doping dependence of $q_{CO}$ was consistent with the calculation when a pseudogap was included on the Fermi surface Figure \ref{fig:3}(A). 
Since the pseudogap itself may result from strong correlations stemming from the parent Mott state, these Lindhard calculations imply a similar origin for the CO. However, as we discuss below, the full 2D static Lindhard calculation shows much stronger intensity along a direction $45^\circ$ away from the static charge order direction Figure \ref{fig:3}(B), casting doubt on the appropriateness of the Lindhard description. 

The second study focused on $T'$-Nd$_2$CuO$_4$ thin films with low doping, extending EI-RXS to map the entire $q_x$-$q_y$ plane. Those measurements revealed a quasi-circular ring-like feature, i.e., a nearly constant-radius in-plane pattern \cite{Kang2019}, Figure \ref{fig:3}(C). Based on a static Lindhard calculation, Figure \ref{fig:3}(D,E), those results were attributed to low-energy glassy charge modulations that reflect the Fermi surface topology. However, experimentally, EI-RXS could not resolve the energy structure of the scattering, leaving the origin the quasi-circular feature, elastic or inelastic, undetermined.

These two studies lead an important consideration: Given the well-known Fermi surface in Bi-2212 and the results of the static Lindhard calculations, if a quasi-circular feature were to exist in Bi-2212, it would not originate from a Fermi surface instability, necessitating an alternative explanation. Indeed, RIXS measurements of Bi-2212 found a quasi-circular pattern in the $q_x$-$q_y$ plane at finite inelastic energies, with the same wavevector magnitude, $|\mathbf{q}|=q_{CO}$ as the static CO peak, $\mathbf{q}_{CO}$ \cite{boschini2021dynamic, Scott_SciAdv_2023}, Figures \ref{fig:3}(F) and \ref{fig:4}. 
These dynamic correlations with the CO wavelength along all directions in the CuO$_2$ plane were termed quasi-circular dynamic correlations (QCDCs). In the original experiment, the RIXS scattering pattern was acquired using an instrument with a medium energy resolution of approximately $800$\,meV \cite{boschini2021dynamic}, Figure \ref{fig:3}(F). 
The results suggested the presence of QCDCs that are broad over the mid-infrared range, approximately from $100$ to $900$\,meV. Furthermore, since static probes like STM have never observed quasi-circles in the scattering patterns of Bi-2212 \cite{Hoffman_2002, Vershinin_2004, Howald_2003, daSilvaNeto_CDW_2014, Yazdani_ARCMP, Comin_review}, the existence of a quasi-circle would imply that it is necessarily dynamic. Later, high energy resolution RIXS measurements (approximately $37$\,meV) tracking the $q_x$-$q_y$ profile of the bond-stretching phonon softening concluded that a similar pattern exists at low energies also (below $70$\,meV), Figure \ref{fig:4}. In Sec.\,\ref{ssec:QCDC_SM} we discuss the phonon-tracking method and the connection between low-energy QCDCs to strange metal behavior \cite{Scott_SciAdv_2023}. Before that, however, we discuss how the quasi-circular in-plane pattern provides insights into the origin of charge order and the effective Coulomb interaction \cite{boschini2021dynamic}. 

Given the quasi-circular pattern observed in the experiments, it is natural to ask what many-body description might produce it. Systematically going through different textbook approaches to calculating the charge susceptibility, it was first noted that a static Lindhard calculation strongly deviated from the observed quasi-circular pattern, Figure \ref{fig:3}(G). Next, to account for the finite energy resolution of the RIXS instruments, the experiments were compared to the dynamic Lindhard susceptibility integrated over an energy window comparable to the experimental resolution. Despite this adjustment, a strong qualitative disagreement persisted. It was further noted that the disagreement remained regardless of the inclusion of a pseudogap in the calculation.

Given the failure of the Lindhard (polarizability) function, directly reflecting the band structure geometry, to capture the most salient features of the experiment, the effective Coulomb interaction was considered. 
Within the random phase approximation (RPA) formalism, features in the susceptibility may emerge from either peaks in the Lindhard (polarizability) function or from minima in the effective Coulomb interaction. In the extreme case of a featureless polarizability function, as indicated by momentum-resolved electron energy loss spectrosocopy (MEELS) measurements of Bi-2212 \cite{Mitrano_PNAS, Hussain_PRX_2019}, the RPA susceptibility is dominated by the form of the effective Coulomb interaction, $V_{\textrm{eff}}$. Using known expressions and experimentally determined parameters \cite{Drozdov_2018, Hwang_2007,Takayanagi_2002}, a $V_{\textrm{eff}}$ was calculated, incorporating both a short-range Hubbard-like component and a long-range potential derived from Poisson’s equation \cite{Becca_1996, Seibold2000, Caprara2017, Kivelson_RMP}. While separately the short-range and long-range components are monotonic functions of $|\mathbf{q}|$, with the former increasing and the latter decreasing, together they form a minimum at an intermediate value inside the scattering Brillouin zone. Remarkably, the minima of $V_{\textrm{eff}}$, Figure \ref{fig:3}(H), form a quasi-circular pattern with a radius and shape that closely match the experimentally observed QCDCs \cite{boschini2021dynamic}. In other words, the form of the Coulomb interaction described above captured the most salient aspect of the RIXS mapping---a quasi-circular scattering structure. It also indicated that the same long-range Coulomb interactions necessary to avoid phase separation were key to the appearance of quasi-circular minima in $V_{\textrm{eff}}$. 

A more detailed theoretical model that simultaneously reproduces various aspects of the RIXS on Bi-2212, similar to the one for NCCO, is still being developed. Nevertheless, following the first reports of QCDCs, theoretical works have also observed similar quasi-circular structures, using the same model that reproduced the RIXS measured plasmons in cuprates \cite{Yamase_PRB_2021, Bejas_ringlike_2022}. Interestingly, these studies converge on an important conceptual point: the QCDCs are likely the consequence of an effective Coulomb potential that must include long-range interactions.

\subsection{Low-energy QCDCs as mediators of strange metal behavior \label{ssec:QCDC_SM}}

\begin{figure}[!htpb]
\begin{center}
\includegraphics[width=0.99\linewidth]{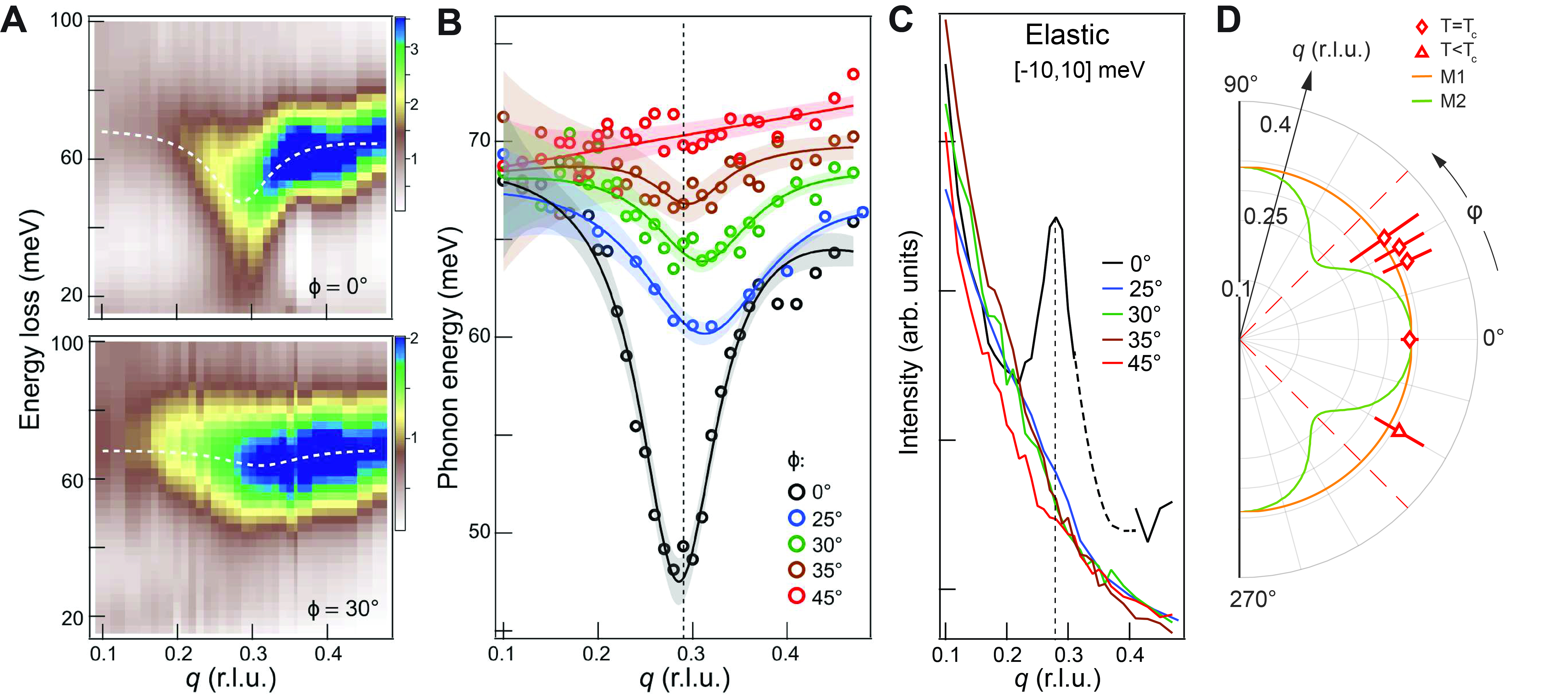}
\end{center}
\caption{
\footnotesize
Low-energy QCDCs measured using the phonon-tracking method. (\textbf{A}) Energy-momentum structure of the RIXS cross-section at $\phi=0^\circ$ and $30^\circ$ after subtraction of the elastic line. (\textbf{B}) Location of the phonon peak obtained by fitting the RIXS spectra for different $\phi$. The solid lines are obtained by fitting the q-dependence of the phonon peak (circles) with a negative Lorentzian function plus a linear background. The shaded regions around the solid lines are generated from the 95\% confidence interval obtained for the various fits to the spectra. The solid lines for $\phi=0^\circ$ and $30^\circ$ in (\textbf{B}) appear as dashed white lines in (\textbf{A}).
(\textbf{C}) Momentum-dependence of the RIXS intensity integrated around the elastic line [$-10,10$] meV as a function of $\phi$.
(\textbf{D}) Polar plot contrasting M1 (diffusive scattering) and M2 (QCDCs) models (orange and green solid lines) and the experimental data (red symbols). The error bars in (\textbf{D}) are obtained from the fits to the phonon dispersion in Fig.\,(\textbf{B}). 
r.l.u., reciprocal lattice units. 
From (Scott et al., 2023) \cite{Scott_SciAdv_2023}. Reprinted with permission from AAAS.
 }\label{fig:4}
\end{figure}

Recent combined transport and RIXS studies revealed an unexpected link between linear-in-temperature resistivity, which characterizes the strange metal behavior, and charge order (CO) in YBCO \cite{Wahlberg_Science_2021}. 
The linear-in-temperature resistivity \cite{T_linear_cuprates_1987, T_linear_BSCO_1989}, often associated with an isotropic scattering rate that depends solely on temperature (i.e., $\propto k_B T / \hbar$, sometimes called the Planckian regime) \cite{grissonnanche2021linear, Varma_MFL, Varma_critical, Patel_PRX, Patel_PRL, Phillips_Science_Review, Patel_Sachdev_StrangeMetal_Science_2023}, has been suggested to result from low-energy dynamic CO leading to effective isotropic scattering \cite{seibold2021strange, caprara2022dissipation}.
In this scenario, isotropic scattering occurs if collective modes, (\textit{i}) low in energy and (\textit{ii}) broad in momentum-space, are available to scatter electrons between any two points on the Fermi surface. 
Naturally, QCDCs occupying such a large momentum space manifold satisfy the second criterion. Demonstrating the first criterion, that they also exist at low energies, required a new approach based on a \textit{phonon-tracking} methodology \cite{Scott_SciAdv_2023}. This is based on the phenomenology revealed by RIXS measurements on various cuprates, which indicate a softening of the RIXS-measured bond-stretching (BS) phonon peak at $\mathbf{q}_{CO}$, i.e., along $q_x$ and $q_y$ \cite{Chaix2017,lee2021melting,Li_PNAS_2020,Miao_PRX_2019,Lin_PRL_2020,Peng_PRL_2020,Huang_PRX_2021, Wang_SciAdv_2021}. By tracking the softening of the RIXS-measured BS phonon in the $q_x$-$q_y$ plane, QCDCs were shown to exist at energies below approximately $70$\,meV \cite{Scott_SciAdv_2023}, Figure \ref{fig:4}(A-B,D). Interestingly, the softening at azimuthal angles close to $45^\circ$ contrasts with the absence of a static CO peak on the elastic line for azimuthal angles away from $0^\circ$, Figure \ref{fig:4}(C). Overall, these measurements demonstrated the dynamic, low-energy nature of the QCDCs, positioning them as a viable candidate to mediate the isotropic scattering in the strange metal phase.

\section{Open questions and future directions}

\subsection{Enabling technical advancements  \label{ssec:developments}}
{Before addressing open questions and future directions, we briefly discuss the key instruments used to obtain the EI-RXS and RIXS results reviewed above, pointing out their key technical characteristics. 
This section is not intended as a comprehensive review of technical developments, but rather a brief mention of the beamlines from which these studies originated. For a more detailed and recent review of RIXS, we refer readers to \cite{NatRev_RIXS_review}.}

{
First, in the studies of charge order using EI-RXS in the soft x-rays, the UE-46 beamline at the BESSY II synchrotron \cite{weschke2018ue46_instrument} and the REIXS beamline at the Canadian Light Source \cite{REIXS_instrument} were instrumental. At REIXS, measurements were conducted down to $20$\,K, with the capability to heat the sample stage up to $400$\,K. The ability to reach high temperatures was crucial for the temperature-dependent studies of electron-doped cuprate \cite{daSilvaNeto_Science_2015, daSilvaNeto_SciAdv_2016}, such as the one presented in Fig.\,\ref{fig:1}(B). Complementarily, the UE-46 beamline allowed access to lower temperatures, down to $10$\,K, and a second endstation at UE-46 enabled the application of magnetic fields at similar temperatures. These features were pivotal for studying the relationship between charge order and superconductivity \cite{Santi_2014}, particularly in low-$T_c$ samples, such as electron-doped cuprates \cite{daSilvaNeto_SciAdv_2016}.
}

{
Second, the evolution of RIXS, as reflected in the results presented above, was driven by advancements in energy resolution at the Cu-$L_3$ edge. Before the developments at the ID32 beamline of the European Synchrotron Radiation Facility, which achieved better than $35$\,meV resolution \cite{BROOKES_ID32_instrument}, the best energy resolution of $130$\,meV was achieved at the ADRESS beamline of the Swiss Light Source \cite{SAXES_instrument, Strocov:bf5029_ADDRESS_instrument} (e.g. \cite{LeTacon2011}). Following this significant breakthrough, similar energy resolutions were achieved at the SIX beamline at the National Synchrotron Light Source II at Brookhaven National Laboratory \cite{2ID_instrument, SIX_instrument} and the I21 beamline at the Diamond Light Source \cite{Zhou:rv5159}. Additionally, the development of pol-RIXS capabilities at ID32 was crucial for the studies on NCCO shown in Fig.\,\ref{fig:1}(F). Future application of pol-RIXS are discussed in the next section.
}

{
Finally, despite the advancements in energy resolution for Cu-$L_3$ RIXS, achieving these high resolutions comes at the cost of reduced flux, leading to longer measurement times. For some applications, however, a lower energy resolution may suffice for initial measurements. For instance, the first detection of QCDCs \cite{boschini2021dynamic} was achieved with an energy resolution of $0.8$\,eV at the qRIXS beamline of the Advanced Light Source at Lawrence Berkeley National Laboratory \cite{CHUANG2022146897_qRIXS_instrument}. The balance between energy resolution and flux there allowed for a full in-plane mapping within 12 to 24 hours, whereas at high-resolution beamlines, this time frame typically applies to q-mapping at a single $\phi$. Thus, there remains significant potential for discovery with instruments operating at resolutions on the order of hundreds of meV.
}

\subsection{Decomposing the high energy CO correlations with polarimetric RIXS}

One of the main advantages of the soft x-ray RIXS cross-section is its sensitivity to various degrees of freedom \cite{Ament_RevModPhys.83.705,Devereaux_PhysRevX.6.041019}, but this also presents complications. Despite the success of the \textit{phonon-tracking} method, the energy profile of the QCDCs, i.e., their energy spectrum, has not been resolved. Thus, although high-energy and low-energy QCDCs appear at the same momenta, it remains possible that they have different origins. Resolving this question requires the ability to decompose the rich RIXS spectrum into its different components. For this, pol-RIXS may be the most effective technique to determine the contributions to the MIR energy scale, as demonstrated in NCCO.
However, these experiments are extremely challenging. The most mature state-of-the-art polarimeter, available at the ID32 beamline of the European Synchrotron Radiation Facility, uses a mirror with an average reflection coefficient of about $0.1$ \cite{BROOKES_ID32_instrument, Braicovich}. Recently, pol-RIXS is also being developed at the the I21 beamline at Diamond Light Source, with a similar average reflection coefficient at the Ni-$L_3$ edge already demonstrated \cite{Diamond_polaimeter}. As a result of the low reflection coefficient, polarimetric RIXS requires acquisition times over ten times longer than standard RIXS to achieve comparable statistics. To fit these measurements within a typical five-day beam time, either energy resolution is sacrificed to allow for momentum mapping, as in NCCO \cite{daSilvaNeto_PRB_2018}, or momentum mapping is abandoned to obtain high-resolution polarimetric spectra, as demonstrated in the resolution of the 2-phonon mode \cite{Scott_PRB_2024}. Furthermore, the actual spectral decomposition is not performed directly by the instrument but requires nuanced data processing. 
Still, there are plenty of cuprate families, doping values, electronic phases and regions of momentum space that remain to be explored with pol-RIXS.
Therefore, we believe that one of the most significant future directions for advancements in soft x-ray RIXS instruments will be the development of fast and reliable polarimetry capabilities.

\subsection{QCDCs in other cuprates and strain} 

It is still unclear if QCDCs exist in other cuprate families. One approach to investigate this would be to use the phonon tracking method to map the $q_x$-$q_y$ plane. However, this method may be affected by varying degrees of coupling between QCDCs and the bond-stretching phonon across different families and it would not probe the high-energy sector.
Following a Brazovskii framework \cite{boschini2021dynamic}, 
the effective Coulomb interaction allows charge order domains to appear along any in-plane direction, while the lattice structure and phonon spectrum determine the direction of static charge order. Thus, manipulating the lattice, for example with uniaxial strain, should compress spectral weight from the QCDC manifold into the more localized static CO peaks. This differential measurement approach could be used to detect QCDCs in other cuprates.
On the other hand, a recent study combining uniaxial strain and RIXS on \lsco{} finds that, while the relative intensity of static CO peaks along $q_x$ and $q_y$ is sensitive to uniaxial strain, the energy and intensity of dynamic CO correlations, as well as the softening of the BS mode, remain equal along $q_x$ and $q_y$ \cite{Martinelly_Dynamic_Static_LSCO_strain}. This latest result in \lsco{} further indicates a non-trivial relation between static and dynamic CO correlations, similar to the findings in \cite{Scott_SciAdv_2023}, which highlight different behaviors of static and dynamic CO correlations with $\phi$, Figure \ref{fig:4}.

Despite the recent successful examples of integrating soft x-ray RXS with uniaxial strain \cite{Boyle_PRR_2021_strain, HH_Kim_PRL_2021_strain, Wang_NatComm_2022_strain, Gupta_Hawthorn_PRB_2023_strain, Martinelly_Dynamic_Static_LSCO_strain}, we note that these measurements are extremely challenging for several reasons: (\textit{i}) technical challenges associated with the integration of the strain devices into the ultra-high-vacuum chambers of the escisting RIXS setups, (\textit{ii}) a high-rate of broken samples upon stress application and (\textit{iii}) difficulty in precisely determining the effective strain on the sample during the RIXS study.  
Barring advances in methods for applying uniaxial strain, more efficient RIXS would greatly benefit these studies by allowing for faster cycling of samples. {Relatedly, epitaxial strain \cite{Bluschke_2018} and chemical pressure \cite{Ruiz2022} has been shown to have significant impact on the out-of-plane coupling of the CO in YBCO. How this translates to the in-plane structure of the dynamic charge correlations remains an open question.}

\subsection{New experimental probes of high energy CO correlations}

In recent years, the advent of free-electron lasers (FELs) has enabled the extension of EI-RXS and RIXS into the time domain, providing new insights into the light-induced dynamics and melting of CO in cuprates \cite{mitrano2019ultrafast,mitrano2020probing,bluschke2024orbital}. In particular, two recent time-resolved EI-RXS studies have reported the dynamic competition between superconductivity and CO in YBCO \cite{wandel2022enhanced,jang2022characterization}. Both studies show that quenching the superconducting phase with near-infrared (near-IR) light results in a transient non-thermal enhancement of the static CO coherence length and peak intensity. These results provided evidence that superconductivity is intimately intertwined with CO, disrupting its spatial coherence \cite{wandel2022enhanced}, and that the light-driven non-thermal state of cuprates shares close similarity with that reached under the magnetic field \cite{jang2022characterization}.
Regarding the response of CO above $T_c$, it has been shown that the near-IR light excitation further suppresses the CO peak intensity \cite{wandel2022enhanced}. Therefore, while these pioneering studies demonstrate that studying the transient evolution of CO correlations as an exciting new direction, they could not resolve the possibly different behaviors of the static and dynamic contributions at $q_{CO}$. To gain access into the transient evolution of dynamic CO correlations, time-resolved RIXS (TR-RIXS) becomes necessary.

Building up on the recent report of the ultrafast renormalization of the on-site Coulomb interaction in cuprates via time-resolved X-ray absorption \cite{Mitrano_PRX_2022_TR_XAS}, we highlight the potential of TR-RIXS to investigate $V_{\textrm{eff}}$ by noting that visible/near-IR light excitations in the normal state of cuprates can transiently change the screening via the generation of hot carriers. Consequently, a change in the screening should affect the local minimum of the effective Coulomb interaction $V_\text{eff}$ discussed in Sec.\,\ref{subsec_HE_QCDCs}, thus potentially transiently altering the QCDCs manifold in the $q_x$-$q_y$ plane -- this could be detected with TR-RIXS. 
And while the current energy resolution of the time-resolved RIXS endstations at the Cu-$L_3$ edge ($\sim$100\,meV) may not allow tracking the transient evolution of the phonon softening described in Sec.\,\ref{ssec:QCDC_SM}, it is certainly sufficient to separate the quasi-static CO peak from the high-energy CO correlations and other underlying collective excitations. 
Furthermore, the proposed studies would benefit from complementary time- and angle-resolved photoemission spectroscopy (TR-ARPES) experiments. TR-ARPES, which can access light-induced changes in the low-energy electronic band structure with momentum resolution \cite{boschini2024time}, allows the simultaneous measurement of transient changes in the band structure. A combined TR-RIXS and TR-ARPES study would thus allow us to assess the effects of the Lindhard polarizability on the transient response of CO correlations, ultimately determining how much it may contribute to the RIXS signal and/or the emergence of QCDCs.
  
The experimental strategy discussed above relies on the combination of two experimental probes to obtain direct insights into the momentum transfer $\textbf{q}$ (RIXS) and the electronic state momentum $\textbf{k}$ (ARPES) with energy resolution. However, by combining these two momentum-resolved techniques, we would still not be able to identify exactly which electronic states $\textbf{k}$ are correlated by $q_{CO}$. In this context, the recently proposed development of two-electron coincidence techniques, such as 2e-ARPES \cite{trutzschler2017band,mahmood2022distinguishing} and noise correlation ARPES \cite{stahl2019noise,su2020coincidence,devereaux2023angle}, may provide direct access to the static and dynamic correlations between two electronic states for the first time. Looking ahead, the successful demonstration and development of these techniques would offer a direct way to quantitatively estimate which electronic states play a role in the emergence of charge correlations in cuprates and in quantum materials in general.

\newpage

\section*{Conflict of Interest Statement}

The authors declare that the research was conducted in the absence of any commercial or financial relationships that could be construed as a potential conflict of interest.

\section*{Author Contributions}

EHdSN: writing – original draft, writing – review \& editing. AF: writing – original draft, writing – review \& editing. FB: writing – original draft, writing – review \& editing.

\section*{Funding}
E.H.d.S.N was supported by the Alfred P. Sloan Fellowship and the National Science Foundation under Grant No. DMR-2034345.
{A.F. was supported by the CIFAR Azrieli Global Scholars program and by the National Science Foundation under Grant No. DMR-2145080.}
{F.B. was supported by the Natural Sciences and Engineering Research Council of Canada,  the Canada Research Chairs Program, the Fonds de recherche du Qu\'{e}bec – Nature et Technologies, and the Minist\`{e}re de l'\'{E}conomie, de l'Innovation et de l'\'{E}nergie - Qu\'{e}bec.}

\section*{Acknowledgments}
We thank Kirsty Scott, Matteo Minola and Yu He for the careful reading of the manuscript and insightful comments.

\bibliographystyle{apsrev4-2}
\bibliography{biblib, rings}

\end{document}